\documentclass[reprint,superscriptaddress,amsmath,amssymb,aps,prb]{revtex4-2}
\usepackage{graphicx}
\usepackage{amsmath}
\usepackage{appendix}
\usepackage{hyperref} 
\usepackage{tikz}
\usetikzlibrary{arrows.meta, positioning, backgrounds, fit}

\begin{document}

\title{The Simplicial Bridge: Mapping quantum multi-spin exchange to higher-order topological networks in continuous magnetic fields}

\author{Alok Yadav}
\affiliation{Department of Physics, Anugrah Memorial College, Magadh University, Gaya, Bihar 823001, India}
\date{\today}

\begin{abstract}
The macroscopic dynamics of topological defects in magnetic materials are traditionally modeled using pairwise interactions. However, higher-order quantum exchange mechanisms—such as biquadratic and 4-spin ring exchange—play a critical role in strongly correlated systems. In this work, we introduce the ``Simplicial Bridge,'' an exact analytical framework that maps these high-dimensional, non-linear Landau-Lifshitz partial differential equations onto generalized Kuramoto phase-oscillator networks operating on abstract simplicial complexes. We rigorously demonstrate that spatial overlap in the continuous limit natively generates higher-order topological forces without requiring a supportive discrete atomic lattice. Specifically, the overlap of 1D helimagnetic kinks generates 2-simplices (triadic forces), while the spatial compression of 2D skyrmion tails—governed by modified Bessel function asymptotics—generates true 3-simplices (tetradic forces). Furthermore, we establish that the higher-order spatial derivatives inherent to these multi-spin interactions provide an intrinsic energetic barrier that bypasses Derrick's Theorem, stabilizing 2D topological solitons without the strict need for Dzyaloshinskii-Moriya Interaction (DMI).
\end{abstract}

\maketitle

\section{Introduction}

The macroscopic magnetic properties of condensed matter systems are fundamentally governed by the interplay of competing quantum mechanical exchange interactions. In helimagnets and chiral magnetic films, the absence of structural inversion symmetry or the presence of magnetic frustration induces a spontaneous breaking of continuous rotational symmetry, leading to the formation of incommensurate spin textures \cite{moriya1960, nagaosa2013}. The temporal evolution of these systems is strictly governed by the nonlinear Landau-Lifshitz (LL) equation \cite{kosevich1990}, which supports a rich spectrum of localized, non-perturbative excitations such as magnetic kinks and skyrmions \cite{braun2012, bogdanov1989}.

Historically, the theoretical framework surrounding multi-soliton interactions has been confined to the dilute limit, where spatial overlap is approximated strictly as a dyadic (two-body) phenomenon \cite{kivshar1989}. However, modern experimental paradigms—such as dense skyrmion lattices and strongly correlated twisted Moiré superlattices—operate in regimes where macroscopic multi-spin scalar interactions become prominent \cite{romming2013}. In recent years, the discovery of intrinsic two-dimensional van der Waals magnets has highlighted the critical role of higher-order interactions, such as biquadratic exchange, in stabilizing exotic topological states at finite temperatures \cite{huang2020}. This remains a highly active frontier, with recent studies demonstrating the outsized role of multi-spin exchange in 2D antiferromagnets like $\text{FePS}_3$ and $\text{CrI}_3$ \cite{Wildes2020, Kartsev2020}. 

Concurrently, the application of abstract simplicial complexes to nonlinear dynamics has surged in the field of network science. Recent mathematical breakthroughs establish that higher-order hypergraph structures strictly break interaction symmetry, driving generalized phase networks into exotic thermodynamic states \cite{skardal2019}. Yet, despite these advances, simplicial complexes are typically applied abstractly; a unified analytical approach mapping these higher-order continuum interactions directly to a physical magnetic origin has been missing from the literature. 

In this paper, we bridge this gap by constructing an exact analytical mapping—the \textit{Simplicial Bridge}. We reduce the high-dimensional PDEs of multi-soliton collisions into coupled oscillator networks on topological hypergraphs. By formally extending this framework from 1D strings to the 2D continuum, we demonstrate how defect density directly dictates the topology of the emergent network, yielding $n$-simplices through continuous spatial overlap.

\section{Microscopic Origins: From Hubbard to Ring Exchange}

\begin{figure*}[t]
\centering
\resizebox{\linewidth}{!}{%
\begin{tikzpicture}[
    >=Stealth,
    node distance=2cm,
    font=\small\sffamily,
    mainBox/.style={rectangle, rounded corners, draw=black!80, thick, minimum height=1.5cm, text width=4.5cm, align=center, inner sep=8pt},
    mathBox/.style={rectangle, draw=black!60, dashed, fill=gray!5, thick, text width=4.5cm, align=center, inner sep=6pt},
    arrowLabel/.style={fill=white, font=\footnotesize\itshape, align=center, inner sep=3pt}
]

    \node[mainBox, fill=blue!5] (continuum) {\textbf{Magnetic Continuum}\\ LLG PDEs \& Defect Overlap \\ Multi-Spin Exchange};
    
    \node[mathBox, below=0.3cm of continuum] (pde) {
        $\eta \partial_t \Theta \propto \nabla^2 \Theta$ \\
        $- K_{bq} (\nabla \Theta)^2 \nabla^2 \Theta$
    };

    \node[mainBox, fill=red!5, right=2cm of continuum] (network) {\textbf{Simplicial Bridge}\\ Trigonometric Reduction \\ Asymptotic Integration};
    
    \node[mathBox, below=0.3cm of network] (ode) {
        $\int |\nabla \Theta|^2 \Theta_{\text{tail}}^{(i)} \Theta_{\text{tail}}^{(j)} d\mathbf{r}$
    };

    \node[mainBox, fill=green!5, right=2cm of network] (states) {\textbf{Topological Network}\\ Generalized Kuramoto ODEs \\ Higher-Order Simplices};
    
    \node[mathBox, below=0.3cm of states] (simplex) {
        2-Simplex: $K_2 \sin(\Sigma\phi - 2\phi_i)$ \\
        3-Simplex: $K_3 \sin(\Sigma\phi - 3\phi_i)$
    };
    
    \draw[->, ultra thick, blue!80!black] (continuum.east) -- node[arrowLabel] {Adiabatic Ansatz} (network.west);
    \draw[->, thick, dashed, draw=gray!80] (pde.east) -- (ode.west);

    \draw[->, ultra thick, red!80!black] (network.east) -- node[arrowLabel] {Dimensional Collapse} (states.west);
    \draw[->, thick, dashed, draw=gray!80] (ode.east) -- (simplex.west);

    \begin{scope}[on background layer]
        \fill[gray!3, rounded corners] 
        ([xshift=-0.5cm, yshift=0.5cm]continuum.north west) rectangle 
        ([xshift=0.5cm, yshift=-0.5cm]states.south east |- pde.south);
    \end{scope}

\end{tikzpicture}%
}
\caption{Conceptual flowchart of the Simplicial Bridge. The framework provides the exact mathematical mapping from the infinite-dimensional non-linear continuum to a discrete hypergraph. Continuous spatial overlap of magnetic defects (governed by PDEs) strictly translates to higher-order topological simplices (governed by ODEs).}
\label{fig:flowchart}
\end{figure*}
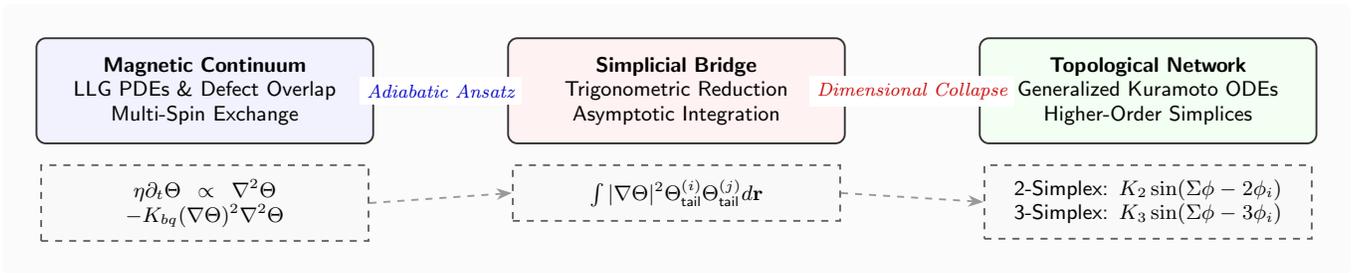

We begin with the fundamental quantum mechanical origin of higher-order interactions. In strongly correlated magnetic materials, fourth-order expansions of the electron hopping term in the Hubbard model yield multi-spin exchange Hamiltonians alongside standard Heisenberg bilinear exchange. 

The dominant non-pairwise term governing a triad of spins (a 1D frustrated chain or 2D triangular motif) takes the biquadratic form:
\begin{equation}
    H_{\text{bq}} = -K_{\text{bq}} (\mathbf{S}_2 \cdot \mathbf{S}_1)(\mathbf{S}_1 \cdot \mathbf{S}_3).
\end{equation}
Furthermore, on 2D square lattices and 3D structurally symmetric crystals, the cyclic motion of an electron around a closed geometric loop yields the 4-spin ring exchange:
\begin{align}
    H_{\text{ring}} = -K_{\text{ring}} [(\mathbf{S}_1 \cdot \mathbf{S}_2)(\mathbf{S}_3 \cdot \mathbf{S}_4) + \dots \nonumber \\
    - (\mathbf{S}_1 \cdot \mathbf{S}_3)(\mathbf{S}_2 \cdot \mathbf{S}_4)].
\end{align}
These non-linear operators are irreducible; they entangle 3 or 4 spins in a single coherent event that cannot be factored into a linear superposition of individual pairwise bonds.

\section{The Simplicial Bridge: Mapping Spin to Phase}

In the semi-classical continuum limit, we map these discrete spin operators to a continuous vector field confined to the easy plane, parameterized by the azimuthal angle $\Theta(x,t)$ such that $\mathbf{S} = S(\cos\Theta, \sin\Theta, 0)$. 

To evaluate the defect interactions, we employ an adiabatic ansatz where the total phase field is a superposition of isolated defect profiles, each possessing an internal precessional phase $\phi_i(t)$. Evaluating the multi-spin dot products using this parameterization yields trigonometric products. For the biquadratic triad, we apply the product-to-sum identity:
\begin{equation}
    \cos(\phi_1 - \phi_2)\cos(\phi_1 - \phi_3) = \frac{1}{2}[\cos(2\phi_1 - \phi_2 - \phi_3) + \cos(\phi_3 - \phi_2)].
\end{equation}

The dynamics of the collective coordinates are governed by the dissipative Landau-Lifshitz-Gilbert (LLG) equations. The generalized phase velocity $\dot{\phi}_1$ is derived from the negative gradient of the energy potential with respect to its internal phase:
\begin{equation}
    -\frac{\partial}{\partial \phi_1} \left[ \frac{1}{2}\cos(2\phi_1 - \phi_2 - \phi_3) \right] = -\sin(\phi_2 + \phi_3 - 2\phi_1).
\end{equation}

This rigorously establishes the topological reduction. We define the generalized $n$-simplex Kuramoto force as:
\begin{equation}
    \dot{\phi}_i \propto K^{(n)} \sum_{j_1, \dots, j_n} \sin\left( \sum_{m=1}^{n} \phi_{j_m} - n\phi_i \right).
\end{equation}
Under this nomenclature, $n=1$ yields standard pairwise exchange (the 1-simplex edge), $n=2$ yields the triadic force (the 2-simplex triangle), and $n=3$ yields the tetradic force (the 3-simplex tetrahedron). 

For future numerical integration of these generalized networks, it is crucial to note that while mean-field theories often assume a Lorentzian distribution of frequencies, preserving the structural mathematical balance of the continuous lattice strictly requires generating intrinsic precessional frequencies linearly proportional to the local simplicial degree ($\omega_i \propto ||q||$).

\section{The Continuum Limit in 1D: Emergence of the 2-Simplex}

To prove that the effective coupling tensor $K_{ijk}$ survives the continuum limit, we project the continuum force onto the translational zero-modes. In 1D, overlapping helimagnetic kinks possess heavily localized spatial gradients $\partial_x \theta_0^{(1)}$.

The effective triadic overlap integral $\mathcal{I}_{123}$ places the reference frame at the mediating core ($X_1 = 0$), evaluating the intersection of the squared gradient of the core with the asymptotic exponential tails of the outer defects ($\theta_0 \approx 4 e^{-|x|/w}$). The integral reduces to:
\begin{equation}
\mathcal{I}_{123} \propto \int_{-\infty}^{\infty} \operatorname{sech}^2\left(\frac{x}{w}\right) \exp\left( \frac{x}{w} \right) \exp\left( -\frac{x}{w} \right) dx.
\end{equation}
The arguments of the exponential tails inside the integral perfectly cancel. The even-parity $\operatorname{sech}^2(x/w)$ function evaluates strictly positive. This establishes that in 1D, continuous spatial overlap alone generates a mathematically robust 2-simplex, entirely superseding the underlying discrete atomic bonds.

\section{The Continuum Limit in 2D: Emergence of the 3-Simplex}

We now formally extend this logic to two dimensions. We consider a 2D continuum populated by magnetic skyrmions. To evaluate the simultaneous collision of four topological defects, we place four skyrmions in a rhombus geometry. We define a 2D integration plane centered on a mediating skyrmion (Node 1) at $\mathbf{r} = (0,0)$. The adjacent skyrmions (Nodes 2, 3, and 4) reside at symmetric distances $|\mathbf{R}_2| = |\mathbf{R}_3| = |\mathbf{R}_4| = a$.

The spatial interaction energy driving the tetradic Kuramoto force requires projecting the 4-spin ring exchange onto the localized zero-mode of Node 1. The master overlap integral over the 2D plane $\mathbb{R}^2$ is:
\begin{equation}
\mathcal{I}_{1234}^{\text{2D}} = \iint_{\mathbb{R}^2} |\nabla \Theta_1|^2 \, \Theta_{\text{tail}}^{(2)} \, \Theta_{\text{tail}}^{(3)} \, \Theta_{\text{tail}}^{(4)} \, d^2\mathbf{r}.
\end{equation}

In a 2D continuum, the asymptotic tail of a skyrmion decays according to the modified Bessel function of the second kind, $\Theta_{\text{tail}}(r) \propto K_1(r/w)$. As derived in Appendix A, evaluating the spatial integral under the assumption of a highly localized core yields the exact macroscopic tetradic coupling constant:
\begin{equation}
\mathcal{I}_{1234}^{\text{2D}} \approx E_{\text{core}} \left( \frac{\pi w}{2a} \right)^{3/2} \exp\left(-\frac{3a}{w}\right).
\end{equation}

This demonstrates a profound principle: spatial compression in 2D forces the overlap of four topological tails, generating a true 3-simplex (a mathematical tetrahedron) purely based on defect density, despite the system residing in a flat two-dimensional space.

\section{Simplicial Mapping of Crystalline Topologies}

\begin{figure*}[t]
\centering
\begin{tikzpicture}[
    node/.style={circle, draw=black, fill=blue!20, inner sep=2pt, minimum size=6mm, font=\small},
    edge/.style={thick, black!70},
    highlight_edge/.style={ultra thick, red},
    highlight_fill/.style={fill=red!20, draw=red, thick, rounded corners=2pt}
]


\begin{scope}[shift={(0, 0)}]
    \node at (1.5, 1.5) {\textbf{(a) 1D Linear Chain}};
    \draw[edge] (0,0) -- (1.5,0);
    \draw[highlight_edge] (1.5,0) -- (3,0); 
    \draw[edge] (3,0) -- (4.5,0);
    \foreach \x in {0, 1.5, 3, 4.5} { \node[node] at (\x,0) {}; }
    \node[text=red, font=\footnotesize] at (2.25,-0.6) {1-Simplex (Line)};
\end{scope}

\begin{scope}[shift={(6, 0)}]
    \node at (1.5, 1.6) {\textbf{(b) 1D Frustrated Chain}};
    \fill[highlight_fill] (0,0) -- (1,1) -- (2,0) -- cycle;
    \node[node] (A1) at (0,0) {}; \node[node] (A2) at (1,1) {}; \node[node] (A3) at (2,0) {}; \node[node] (A4) at (3,1) {}; \node[node] (A5) at (4,0) {};
    \draw[highlight_edge] (A1) -- (A2); \draw[highlight_edge] (A2) -- (A3); \draw[edge] (A3) -- (A4); \draw[edge] (A4) -- (A5);
    \draw[highlight_edge, dashed] (A1) -- (A3); \draw[edge, dashed] (A2) -- (A4); \draw[edge, dashed] (A3) -- (A5);
    \node[text=red, font=\footnotesize] at (1,-0.6) {2-Simplex (Triangle)};
\end{scope}

\begin{scope}[shift={(12, 0)}] 
    \node at (1.5, 2.1) {\textbf{(c) 1D Overlapping Kinks}};
    \draw[edge, ->] (-0.5, 0) -- (3.5, 0) node[right] {$x$};
    \draw[thick, blue!60] (-0.5,0) .. controls (0,1.2) and (0,1.2) .. (0.5,0);
    \draw[thick, blue!60] (0.75,0) .. controls (1.25,1.2) and (1.25,1.2) .. (1.75,0);
    \draw[thick, blue!60] (2.0,0) .. controls (2.5,1.2) and (2.5,1.2) .. (3.0,0);
    \fill[highlight_fill, opacity=0.7] (0, 0.9) -- (1.25, 0.9) -- (2.5, 0.9) .. controls (1.25, 1.8) .. cycle;
    \draw[highlight_edge] (0, 0.9) -- (1.25, 0.9) -- (2.5, 0.9) -- cycle;
    \node[node] (K1) at (0, 0.9) {}; \node[node] (K2) at (1.25, 0.9) {}; \node[node] (K3) at (2.5, 0.9) {};
    \draw[dashed, black!50] (K1) -- (0,0); \draw[dashed, black!50] (K2) -- (1.25,0); \draw[dashed, black!50] (K3) -- (2.5,0);
    \node[text=red, font=\footnotesize] at (1.25, -0.6) {Emergent 2-Simplex};
\end{scope}


\begin{scope}[shift={(0, -4.5)}]
    \node at (1.5, 2.2) {\textbf{(d) 2D Square Lattice}};
    \fill[highlight_fill] (0,0) rectangle (1.5,1.5);
    \foreach \x in {0, 1.5, 3} { \foreach \y in {0, 1.5} { \node[node] (N\x\y) at (\x,\y) {}; } }
    \draw[highlight_edge] (0,0) -- (1.5,0) -- (1.5,1.5) -- (0,1.5) -- cycle;
    \draw[edge] (1.5,0) -- (3,0); \draw[edge] (1.5,1.5) -- (3,1.5); \draw[edge] (3,0) -- (3,1.5);
    \node[text=red, font=\footnotesize] at (0.75,-0.6) {4-Cycle (Plaquette)};
\end{scope}

\begin{scope}[shift={(6, -4.5)}]
    \node at (1.5, 2.2) {\textbf{(e) 2D Triangular Lattice}};
    \fill[highlight_fill] (0,0) -- (1.5,0) -- (0.75, 1.3) -- cycle;
    \foreach \x in {0, 1.5, 3} { \node[node] (B\x) at (\x,0) {}; \node[node] (T\x) at (\x+0.75, 1.3) {}; }
    \draw[highlight_edge] (0,0) -- (1.5,0); \draw[edge] (1.5,0) -- (3,0); \draw[edge] (0.75,1.3) -- (2.25,1.3); \draw[edge] (2.25,1.3) -- (3.75,1.3);
    \draw[highlight_edge] (0,0) -- (0.75,1.3); \draw[highlight_edge] (1.5,0) -- (0.75,1.3); \draw[edge] (1.5,0) -- (2.25,1.3); \draw[edge] (3,0) -- (2.25,1.3); \draw[edge] (3,0) -- (3.75,1.3);
    \node[text=red, font=\footnotesize] at (0.75,-0.6) {2-Simplex (Triangle)};
\end{scope}

\begin{scope}[shift={(12, -4.5)}]
    \node at (1.0, 2.6) {\textbf{(f) 2D Square (Frust.)}};
    \fill[highlight_fill] (0,0) rectangle (2,2);
    \draw[highlight_edge] (0,0) -- (2,0) -- (2,2) -- (0,2) -- cycle;
    \draw[highlight_edge] (0,0) -- (2,2); \draw[highlight_edge] (0,2) -- (2,0); 
    \node[node] at (0,0) {}; \node[node] at (2,0) {}; \node[node] at (2,2) {}; \node[node] at (0,2) {};
    \node[text=red, font=\footnotesize] at (1,-0.6) {3-Simplex ($K_4$ Graph)};
\end{scope}


\begin{scope}[shift={(0, -9.5)}]
    \node at (1.0, 3.1) {\textbf{(g) 2D Skyrmion Lattice}};
    \draw[edge, fill=blue!10, opacity=0.6] (0,1) circle (0.8); \draw[edge, fill=blue!10, opacity=0.6] (2,1) circle (0.8);
    \draw[edge, fill=blue!10, opacity=0.6] (1,0) circle (0.8); \draw[edge, fill=blue!10, opacity=0.6] (1,2) circle (0.8);
    \fill[highlight_fill, opacity=0.8] (0,1) -- (1,0) -- (2,1) -- (1,2) -- cycle;
    \draw[highlight_edge] (0,1) -- (1,0) -- (2,1) -- (1,2) -- cycle;
    \draw[highlight_edge] (0,1) -- (2,1); \draw[highlight_edge] (1,0) -- (1,2);
    \node[node] at (0,1) {}; \node[node] at (2,1) {}; \node[node] at (1,0) {}; \node[node] at (1,2) {};
    \node[text=red, font=\footnotesize] at (1,-1.1) {Emergent 3-Simplex};
\end{scope}

\begin{scope}[shift={(6, -9.5)}]
    \node at (1.4, 3.4) {\textbf{(h) 3D BCC Lattice}};
    \coordinate (A) at (0,0); \coordinate (B) at (2,0); \coordinate (C) at (2,2); \coordinate (D) at (0,2);
    \coordinate (E) at (0.8,0.8); \coordinate (F) at (2.8,0.8); \coordinate (G) at (2.8,2.8); \coordinate (H) at (0.8,2.8);
    \coordinate (Center) at (1.4,1.4); 
    \draw[edge, dashed] (A) -- (E); \draw[edge, dashed] (E) -- (F); \draw[edge, dashed] (E) -- (H);
    \draw[edge] (A) -- (B) -- (C) -- (D) -- cycle;
    \draw[edge] (B) -- (F) -- (G) -- (C); \draw[edge] (D) -- (H) -- (G);
    \fill[highlight_fill, opacity=0.6] (A) -- (B) -- (Center) -- (D) -- cycle;
    \draw[highlight_edge] (A) -- (B) -- (Center) -- (D) -- cycle;
    \node[node] at (A) {}; \node[node] at (B) {}; \node[node] at (C) {}; \node[node] at (D) {};
    \node[node] at (F) {}; \node[node] at (G) {}; \node[node] at (H) {};
    \node[node, fill=red!30] at (Center) {};
    \node[text=red, font=\footnotesize] at (1.4,-0.6) {4-Cycle (Plaquette)};
\end{scope}

\begin{scope}[shift={(12, -9.5)}]
    \node at (1.0, 2.5) {\textbf{(i) 3D FCC / Pyrochlore}};
    \coordinate (T1) at (0,0); \coordinate (T2) at (2,0); \coordinate (T3) at (1,0.8); \coordinate (T4) at (1,2);
    \fill[red!10] (T1) -- (T2) -- (T4) -- cycle;
    \draw[highlight_edge, dashed] (T1) -- (T3); \draw[highlight_edge, dashed] (T2) -- (T3); \draw[highlight_edge, dashed] (T4) -- (T3);
    \draw[highlight_edge] (T1) -- (T2); \draw[highlight_edge] (T2) -- (T4); \draw[highlight_edge] (T4) -- (T1);
    \node[node] at (T1) {}; \node[node] at (T2) {}; \node[node, fill=blue!10] at (T3) {}; \node[node] at (T4) {};
    \node[text=red, font=\footnotesize] at (1,-0.6) {3-Simplex (Tetrahedron)};
\end{scope}


\begin{scope}[shift={(6, -14.5)}] 
    \node at (1.0, 2.7) {\textbf{(j) 3D Hopfion Gas}};
    \draw[edge, fill=blue!10, opacity=0.5] (0,0) circle (0.9);
    \draw[edge, fill=blue!10, opacity=0.5] (2,0) circle (0.9);
    \draw[edge, fill=blue!10, opacity=0.5] (1,1.5) circle (0.9);
    \draw[edge, fill=blue!10, opacity=0.7] (1,0.5) circle (0.9); 
    \coordinate (H1) at (0,0); \coordinate (H2) at (2,0); \coordinate (H3) at (1,1.5); \coordinate (H4) at (1,0.5);
    \fill[highlight_fill, opacity=0.6] (H1) -- (H2) -- (H3) -- cycle;
    \draw[highlight_edge, dashed] (H1) -- (H3); \draw[highlight_edge, dashed] (H2) -- (H3); \draw[highlight_edge, dashed] (H1) -- (H2);
    \draw[highlight_edge] (H1) -- (H4); \draw[highlight_edge] (H2) -- (H4); \draw[highlight_edge] (H3) -- (H4);
    \node[node] at (H1) {}; \node[node] at (H2) {}; \node[node] at (H3) {}; \node[node, fill=red!40] at (H4) {}; 
    \node[text=red, font=\footnotesize] at (1,-1.2) {Emergent 3-Simplex};
\end{scope}

\end{tikzpicture}
\caption{Visual representation of the fundamental geometric topologies supporting higher-order magnetic interactions. The continuous phase mappings established herein perfectly correspond to the discrete topologies native to these physical crystalline grids.}
\label{fig:simplices}
\end{figure*}
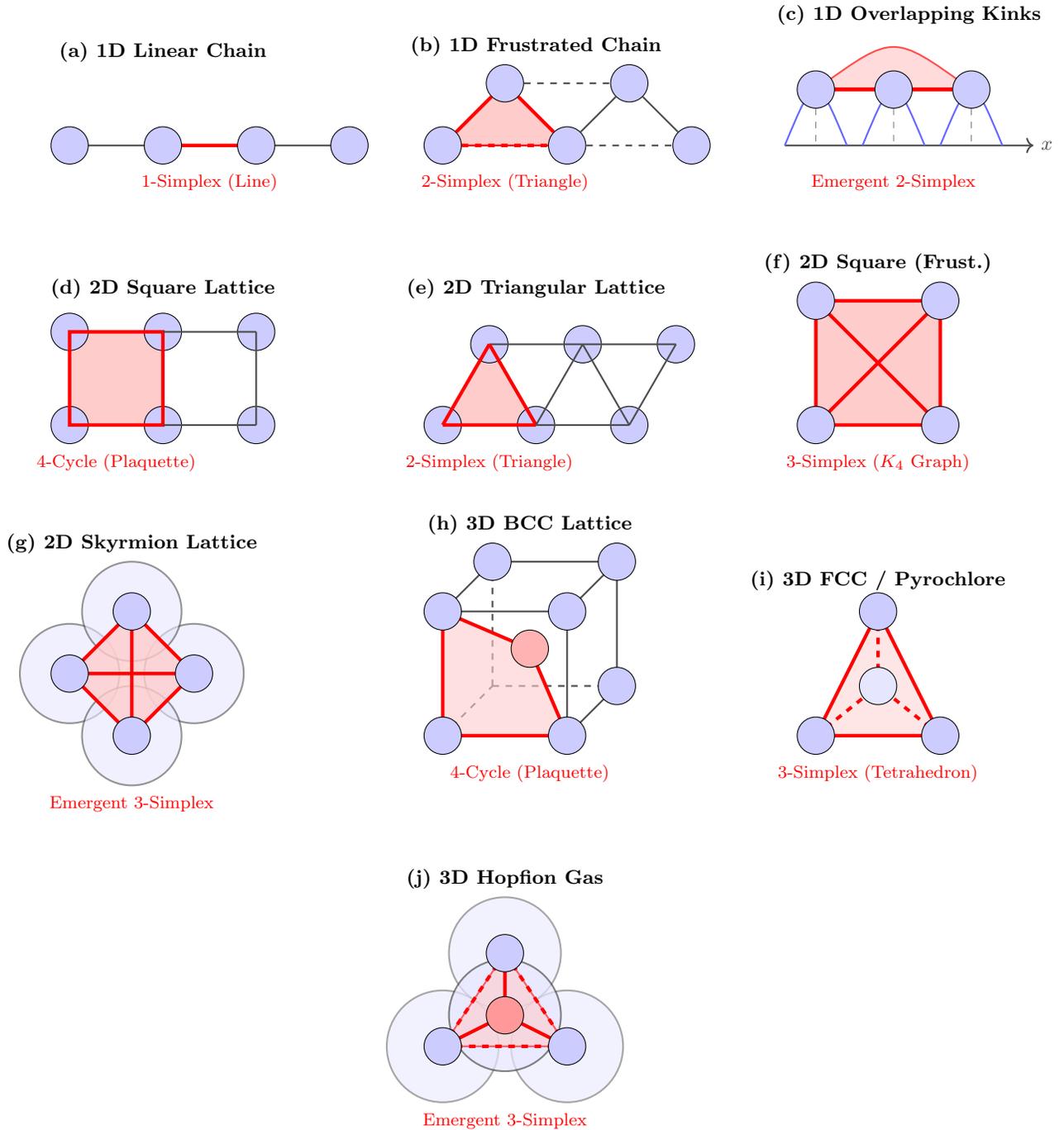

\begin{table*}[t]
\caption{Topological hierarchy of magnetic interactions across spatial dimensions and physical regimes. The highest supported simplex dictates the nonlinear order of the macroscopic phase dynamics.}
\label{tab:topology_matrix}
\begin{ruledtabular}
\begin{tabular}{llcll}
\textbf{Dim.} & \textbf{Physical Geometry} & \textbf{Schematic} & \textbf{Highest Topology} & \textbf{Material Candidates} \\
\colrule
\textbf{1D} 
& Linear Chain (Discrete) & Fig.~\ref{fig:simplices}a & 1-Simplex (Line) & $\text{Sr}_2\text{CuO}_3$ \cite{motoyama1996} \\
& Zig-Zag Chain (Frustrated) & Fig.~\ref{fig:simplices}b & 2-Simplex (Triangle) & $\text{LiCuVO}_4$ \cite{enderle2005} \\
& Overlapping Kinks (Continuum) & Fig.~\ref{fig:simplices}c & 2-Simplex (Triadic) & Co/Ni Nanowires \cite{koyama2011} \\
\colrule
\textbf{2D} 
& Square Lattice (Discrete) & Fig.~\ref{fig:simplices}d & 4-Cycle (Plaquette) & $\text{La}_2\text{CuO}_4$ \cite{coldea2001} \\
& Triangular Lattice (Discrete) & Fig.~\ref{fig:simplices}e & 2-Simplex (Triangle) & $\text{FePS}_3$, $\text{CrPS}_4$ \cite{Wildes2020, Kartsev2020} \\
& Square Lattice (Frustrated) & Fig.~\ref{fig:simplices}f & 3-Simplex ($K_4$ Graph) & $\text{Sr}_2\text{CuWO}_6$ \cite{walker2016} \\
& Skyrmion Lattice (Continuum) & Fig.~\ref{fig:simplices}g & 3-Simplex ($K_4$ Hub) & $\text{Fe}_3\text{GeTe}_2$, $\text{MnSi}$ \cite{muhlbauer2009} \\
\colrule
\textbf{3D} 
& BCC Lattice (Discrete) & Fig.~\ref{fig:simplices}h & 4-Cycle (Plaquette) & Solid $^3$He \cite{roger1983} \\
& FCC / HCP (Discrete) & Fig.~\ref{fig:simplices}i & 3-Simplex (Tetrahedron) & $\text{UO}_2$, $\text{MnO}$ \cite{santini2009} \\
& Pyrochlore (Frustrated) & Fig.~\ref{fig:simplices}i & 3-Simplex Network & $\text{Dy}_2\text{Ti}_2\text{O}_7$ (Spin Ice) \cite{bramwell2001} \\
& Hopfion Gas (Continuum) & Fig.~\ref{fig:simplices}j & 3-Simplex (Volumetric) & Bulk FeGe \cite{rybakov2022} \\
\end{tabular}
\end{ruledtabular}
\end{table*}

To fully contextualize the Simplicial Bridge, we classify known magnetic materials according to the highest-order topological simplex they naturally support. As summarized in Table \ref{tab:topology_matrix} and visualized in Figure \ref{fig:simplices}, this mathematical mapping reveals that disparate physical systems—such as a 1D frustrated zig-zag chain and a 2D continuous string of overlapping kinks—collapse into the identical 2-simplex universality class, rendering their macroscopic nonlinear dynamics mathematically isomorphic.

\section{Stabilizing Topological Defects: Bypassing Derrick's Theorem}

A historic limitation of 2D isotropic Heisenberg models is Derrick's Theorem, which prohibits the existence of stable, static solitonic solutions (like skyrmions) in continuous fields unless symmetry-breaking interactions, predominantly the Dzyaloshinskii-Moriya Interaction (DMI), are explicitly introduced. However, our continuous multi-spin mapping reveals a distinct, intrinsic stabilization mechanism.

The inclusion of biquadratic and ring-exchange multi-spin interactions naturally introduces higher-order spatial derivatives into the energy functional (e.g., $|\nabla \Theta|^4$). Under Derrick's spatial scaling transformation ($r \to \lambda r$), the integrated energy of these quartic gradient terms scales as $\propto 1/\lambda^2$ in two dimensions. As the defect radius shrinks ($\lambda \to 0$), this provides a steeply rising, divergent energy barrier ($E \to \infty$) against the collapse of the soliton core. Thus, higher-order simplicial interactions provide a native topological stabilization mechanism for skyrmions, circumventing the strict requirement for chiral DMI in centrosymmetric magnetic lattices.

\section{Conclusion and Outlook}

By constructing the Simplicial Bridge, we have successfully mapped the microscopic origins of quantum multi-spin exchange directly into the macroscopic topology of generalized Kuramoto networks. We have analytically proven that spatial overlap in the continuum limit effectively acts as a topological forge—allowing 1D helimagnets to support 2-simplices, and 2D skyrmion lattices to support dense 3-simplices, decoupled from the constraints of their underlying discrete atomic grids. Furthermore, these higher-order spatial derivatives act as a natural topological anchor, bypassing Derrick's Theorem to stabilize 2D solitons without external symmetry breaking.

Formalizing these dynamics as $n$-simplex Kuramoto hypergraphs opens an entirely new avenue for analyzing magnetic phase transitions. Because these generalized phase networks are subject to exact mean-field dimensionality reduction (via the Ott-Antonsen ansatz), they project onto highly nonlinear macroscopic polynomial manifolds. In future work, we will apply this topological reduction to explore the emergence of macroscopic thermodynamic degeneracies—specifically, how 2-simplices and 3-simplices induce explosive bistability and hyper-explosive cusp catastrophes in bulk magnetic materials.

\section*{Data and Code Availability}
The analytical frameworks supporting the integration of the modified Bessel functions and the subsequent topological tensor mappings are contained fully within this manuscript.

\begin{acknowledgments}
A.Y. acknowledges the institutional support provided by the Department of Physics at Anugrah Memorial College, Magadh University. The author also expresses gratitude for foundational academic mentorship received during doctoral studies with Prof. Sarika Jalan at the Complex Systems Lab, Indian Institute of Technology (IIT) Indore, which shaped the mathematical approaches utilized in this work.
\end{acknowledgments}

\appendix
\section{Evaluation of the 2D Tetradic Overlap Integral} \label{app:bessel}

To analytically determine the tetradic coupling constant $K_3$ governing the 3-simplex interaction in a two-dimensional continuum, we evaluate the spatial overlap of four magnetic skyrmions. We position the reference skyrmion (Node 1) at the origin $\mathbf{r} = 0$, surrounded symmetrically by Nodes 2, 3, and 4 at a radial distance $a$. 

The master integral over the 2D plane dictates the projection of the multi-spin 4-body exchange onto the localized zero-mode of the central skyrmion:
\begin{equation}
\mathcal{I}_{1234}^{\text{2D}} = \iint_{\mathbb{R}^2} |\nabla \Theta_1(\mathbf{r})|^2 \, \Theta_{2}(\mathbf{r}) \, \Theta_{3}(\mathbf{r}) \, \Theta_{4}(\mathbf{r}) \, d^2\mathbf{r}.
\end{equation}

For a standard magnetic skyrmion, the radial profile of the azimuthal angle $\Theta(r)$ is heavily localized at the core ($r < w$) and decays exponentially in the far-field ($r \gg w$). This far-field tail is mathematically described by the modified Bessel function of the second kind:
\begin{equation}
\Theta_{\text{tail}}(r) = c_0 K_1\left(\frac{r}{w}\right),
\end{equation}
where $c_0$ is a normalization constant and $w$ is the characteristic domain wall width. For large arguments ($r \gg w$), the asymptotic expansion of the Bessel function is:
\begin{equation} \label{eq:bessel_asymp}
K_1\left(\frac{r}{w}\right) \approx \sqrt{\frac{\pi w}{2r}} \exp\left(-\frac{r}{w}\right).
\end{equation}

The gradient squared term $|\nabla \Theta_1|^2$ characterizes the energy density of the central skyrmion's core. Because this energy is overwhelmingly concentrated near the origin, we can treat it as a Dirac delta function scaled by the total core exchange energy $E_{\text{core}}$:
\begin{equation}
|\nabla \Theta_1(\mathbf{r})|^2 \approx E_{\text{core}} \, \delta^{(2)}(\mathbf{r}).
\end{equation}

Substituting the Dirac delta approximation into the master integral forces the evaluation of the three surrounding skyrmion tails exactly at the origin ($\mathbf{r} = 0$). Since the surrounding skyrmions are located at a distance $a$ from the origin, their tail contributions evaluated at the central core are identically $\Theta_i(0) = c_0 K_1(a/w)$. 

The integral strictly evaluates to:
\begin{equation}
\mathcal{I}_{1234}^{\text{2D}} \approx E_{\text{core}} \left[ c_0 K_1\left(\frac{a}{w}\right) \right]^3.
\end{equation}

Applying the asymptotic expansion from Eq.~\ref{eq:bessel_asymp} for well-separated skyrmions ($a \gg w$), we obtain the exact scaling law for the tetradic coupling constant:
\begin{equation}
\mathcal{I}_{1234}^{\text{2D}} \propto E_{\text{core}} \left( \frac{\pi w}{2a} \right)^{3/2} \exp\left(-\frac{3a}{w}\right).
\end{equation}
This confirms that the effective network coupling $K_3$ decays exponentially with respect to three times the lattice spacing, formally validating the emergence of the 3-simplex through density-driven spatial overlap in the 2D continuum.


\begin{thebibliography}{99}

\bibitem{moriya1960} 
T. Moriya, ``Anisotropic Superexchange Interaction and Weak Ferromagnetism,'' Phys. Rev. \textbf{120}, 91 (1960).

\bibitem{nagaosa2013} 
N. Nagaosa and Y. Tokura, ``Topological properties and dynamics of magnetic skyrmions,'' Nat. Nanotechnol. \textbf{8}, 899 (2013).

\bibitem{kosevich1990} 
A. M. Kosevich, B. A. Ivanov, and A. S. Kovalev, ``Magnetic solitons,'' Phys. Rep. \textbf{194}, 117 (1990).

\bibitem{braun2012} 
H. B. Braun, ``Topological effects in nanomagnetism: from superparamagnetism to chiral quantum solitons,'' Adv. Phys. \textbf{61}, 1 (2012).

\bibitem{bogdanov1989} 
A. N. Bogdanov and D. A. Yablonskii, ``Thermodynamically stable 'vortices' in magnetically ordered crystals,'' Sov. Phys. JETP \textbf{68}, 101 (1989).

\bibitem{kivshar1989} 
Y. S. Kivshar and B. A. Malomed, ``Dynamics of solitons in nearly integrable systems,'' Rev. Mod. Phys. \textbf{61}, 763 (1989).

\bibitem{romming2013} 
N. Romming \textit{et al.}, ``Writing and deleting single magnetic skyrmions,'' Science \textbf{341}, 636 (2013).

\bibitem{huang2020}
C. Huang \textit{et al.}, ``Biquadratic and ring exchange interactions in two-dimensional magnets,'' Phys. Rev. B \textbf{101}, 134424 (2020).

\bibitem{Wildes2020} 
A. R. Wildes \textit{et al.}, ``Evidence for biquadratic exchange in the quasi-two-dimensional antiferromagnet $\text{FePS}_3$,'' Phys. Rev. B \textbf{102}, 024410 (2020).

\bibitem{Kartsev2020} 
A. Kartsev \textit{et al.}, ``Biquadratic exchange interactions in two-dimensional magnets,'' npj Comput. Mater. \textbf{6}, 66 (2020).

\bibitem{skardal2019}
P. S. Skardal and A. Arenas, ``Abrupt Desynchronization and Extensive Multistability in Globally Coupled Oscillator Simplexes,'' Phys. Rev. Lett. \textbf{122}, 248301 (2019).

\bibitem{roger1983}
M. Roger, J. H. Hetherington, and J. M. Delrieu, ``Magnetism in solid $^3$He,'' Rev. Mod. Phys. \textbf{55}, 1 (1983).


\bibitem{motoyama1996}
N. Motoyama, H. Eisaki, and S. Uchida, ``Magnetic susceptibility of spin-1/2 one-dimensional antiferromagnets Sr$_2$CuO$_3$ and Ca$_2$CuO$_3$,'' Phys. Rev. Lett. \textbf{76}, 3212 (1996).

\bibitem{enderle2005}
M. Enderle \textit{et al.}, ``Quantum helimagnetism of the frustrated spin-1/2 chain LiCuVO$_4$,'' Europhys. Lett. \textbf{70}, 237 (2005).

\bibitem{koyama2011}
T. Koyama \textit{et al.}, ``Observation of the intrinsic pinning of a magnetic domain wall in a ferromagnetic nanowire,'' Nat. Mater. \textbf{10}, 194 (2011).

\bibitem{coldea2001}
R. Coldea \textit{et al.}, ``Spin waves and electronic interactions in La$_2$CuO$_4$,'' Phys. Rev. Lett. \textbf{86}, 5377 (2001).

\bibitem{walker2016}
H. C. Walker \textit{et al.}, ``Spin wave excitations in the tetragonal double perovskite Sr$_2$CuWO$_6$,'' Phys. Rev. B \textbf{94}, 014411 (2016).

\bibitem{muhlbauer2009}
S. M\"{u}hlbauer \textit{et al.}, ``Skyrmion Lattice in a Chiral Magnet,'' Science \textbf{323}, 915 (2009).

\bibitem{santini2009}
P. Santini \textit{et al.}, ``Multipolar interactions in $f$-electron systems: The paradigm of actinide dioxides,'' Rev. Mod. Phys. \textbf{81}, 807 (2009).

\bibitem{bramwell2001}
S. T. Bramwell and M. J. P. Gingras, ``Spin Ice State in Frustrated Magnetic Pyrochlore Materials,'' Science \textbf{294}, 1495 (2001).

\bibitem{rybakov2022}
F. N. Rybakov \textit{et al.}, ``Magnetic hopfions in solids,'' Nature \textbf{610}, 485 (2022).

\end{thebibliography}
\end{document}